\documentclass[fleqn,10pt]{wlscirep}

\usepackage[]{xcolor}
\definecolor{darkred}{HTML}{990000}
\definecolor{blue}{HTML}{0000FF}

\title{Gene length as a regulator for ribosome recruitment and protein synthesis: theoretical insights}
\author[1,2]{Lucas D. Fernandes}
\author[2]{Alessandro P.S. de Moura}
\author[3,4,*]{Luca Ciandrini}
\affil[1]{Departamento de Entomologia e Acarologia, Escola Superior de Agricultura "Luiz de Queiroz" - Universidade de S\~{a}o Paulo, ESALQ - USP, 13418-900, Piracicaba/SP, Brazil}
\affil[2]{Institute for Complex Systems and Mathematical Biology, University of Aberdeen, Aberdeen, AB24 3UE, UK}
\affil[3]{DIMNP UMR 5235, Universit\'e de Montpellier and CNRS, F-34095, Montpellier, France}
\affil[4]{Laboratoire Charles Coulomb UMR5221, Universit\'e de Montpellier and CNRS, F-34095, Montpellier, France}

\affil[*]{luca.ciandrini@umontpellier.fr}

\begin{abstract}
Protein synthesis rates are determined, at the translational level, by properties of the transcript's sequence. The efficiency of an mRNA can be tuned by varying the ribosome binding sites controlling the recruitment of the ribosomes, or the codon usage establishing the speed of protein elongation.
In this work we propose transcript length as a further key determinant of translation efficiency. Based on a physical model that considers the kinetics of ribosomes advancing on the mRNA and diffusing in its surrounding, as well as mRNA circularisation and ribosome drop-off, we explain how the transcript length may play a central role in establishing ribosome recruitment and the overall translation rate of an mRNA. 
According to our results, the proximity of the 3' end to the ribosomal recruitment site of the mRNA could induce a feedback in the translation process that would favour the recycling of ribosomes.
We also demonstrate how this process may be involved in shaping the experimental ribosome density-gene length dependence. Finally, we argue that cells could exploit this mechanism to adjust and balance the usage of its ribosomal resources.
\end{abstract}

\begin{document}

\flushbottom
\maketitle

\thispagestyle{empty}

\section*{Introduction}
mRNA translation is, together with transcription, the pillar of the central dogma of molecular biology. In spite of its key role in protein synthesis, the accurate understanding of its dynamical details still remains elusive at the present time, and the sequence determinants of mRNA translation efficiency are not fully understood~\cite{gingold_determinants_2011, quax_codon_2015}. Initiation of translation regulates the recruitment of ribosomes and it is believed to be modulated by mRNA secondary structures~\cite{kudla_coding-sequence_2009, salis_automated_2009}, while protein elongation is mainly considered to be regulated by tRNA abundances determining the pace of the ribosome~\cite{plotkin_synonymous_2011, kemp_yeast_2013, gorgoni_identification_2016}. 
The individual steps of translation are thought to be well understood, yet there is no reliable approach quantitatively predicting the overall protein synthesis rates for a given transcript.

A better understanding of the molecular mechanisms of mRNA translation will unravel the physiological determinants of translation efficiency. Besides, this knowledge will be extremely useful in developing applications in synthetic biology and will allow tight control on the average production of a protein and on its expression noise.

\begin{figure}[t!]
\centering
\includegraphics[width=0.7\linewidth]{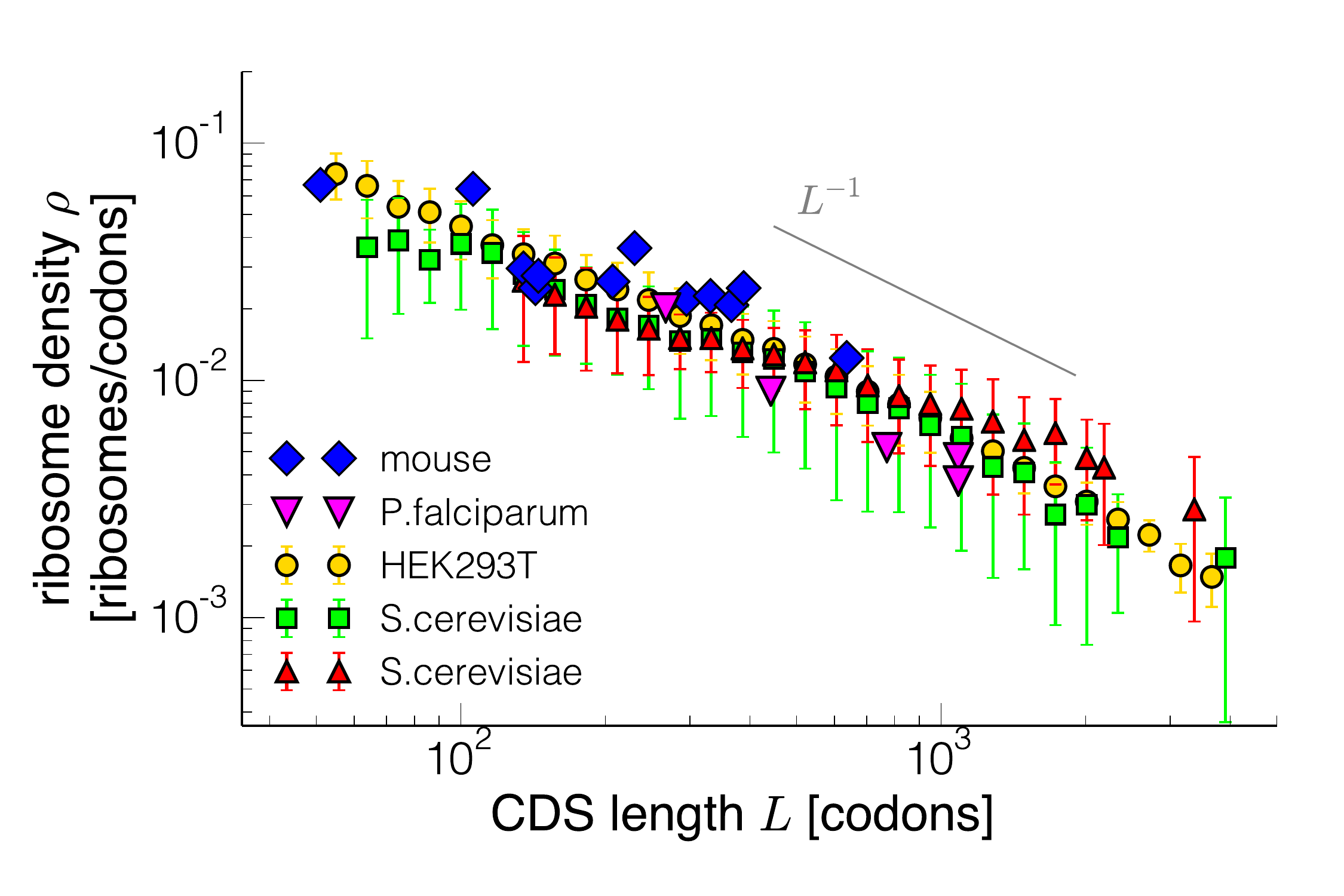}
\caption{Ribosome density vs CDS length for different datasets. Blue diamonds (mice\cite{cataldo_quantitative_1999}) and fuchsia down triangles (\textit{P. falciparum}~\cite{lacsina_polysome_2011}) are individual genes, while yellow circles (HEK293T\cite{hendrickson_concordant_2009}), green squares (\textit{S. cerevisiae}\cite{arava_genome-wide_2003}) and red triangles (\textit{S. cerevisiae}\cite{mackay_gene_2004}) are length-binned data for the entire genome, with the error bars representing the standard deviation for each bin. The grey line indicates the behaviour of a power law with exponent $-1$. \label{fig::1}}
\end{figure}

The translation efficiency of a transcript is often identified with its experimentally observed polysome state (a transcript with two or more ribosomes; monosome when there is only one ribosome residing on it), meaning that transcripts with high ribosome density are more efficiently translated~\cite{li_how_2015}. Remarkably, many experimental observations show that the ribosome density is related to the length $L$ of the coding sequence (CDS): the longer the mRNA, the smaller the ribosomal density. This indicates the presence of a length-dependent control of translation. As we show in Figure~\ref{fig::1}, the observation that average ribosomal densities $\rho$ strongly anti-correlate to CDS lengths $L$ appears to be a conserved feature across many organisms, ranging from unicellular systems such as \textit{L. lactis}~\cite{picard_bacterial_2012}, \textit{S. cerevisiae}~\cite{arava_genome-wide_2003, mackay_gene_2004, ingolia_genome-wide_2009} or \textit{P. falciparum}~\cite{lacsina_polysome_2011}, to more 
complex organisms such as mouse 
and human cells~\cite{cataldo_quantitative_1999, hendrickson_concordant_2009}. The common traits in the density-length dependence suggest that this relationship is dictated by universal mechanisms underlying the translation process.

However, this remark has been strangely overlooked in the literature (with the exception of Guo {\it et al.}~\cite{guo_length-dependent_2015}), particularly in the theoretical literature trying to provide models of mRNA translation. A few hypotheses have been proposed to justify the emergence of the length dependence of the ribosome density, which requires a regulation apparatus acting at the initiation~\cite{arava_compaction_2009} or at the elongation stage~\cite{ingolia_genome-wide_2009}. These hypotheses have not been examined with the support of a mechanistic model and a mathematical approach. In contrast with previous studies, here we explain qualitatively and quantitatively the relationship between ribosome density and CDS length, making the point that transcript length is a critical determinant of protein synthesis rates.

In Figure~\ref{fig::1} we show a log-log plot of measured ribosomal densities as a function of the CDS length for different organisms. The figure suggests a power-law behaviour ($L^{-1}$ is drawn as reference). However, extracting a scaling law from this kind of data is a phenomenological (and probably a too simplistic) description of this relationship that, moreover, can only be measured for a few orders of magnitude in $L$.

Instead, in this paper we propose a mechanistic explanation for the length dependence of translation that is found in experimental data: we describe how the proximity of the mRNA ends increases the local concentration of ribosomes close to their binding sites via a feedback mechanism, favouring their recruitment (and recycling) in short transcripts. In the last part of this work we show how this mechanism could be exploited by the cell to adjust and balance its ribosomal resources.

\section*{Results}


\subsection*{A stochastic model of translation}
\begin{figure}[ht]
\centering
\includegraphics[width=.85\linewidth]{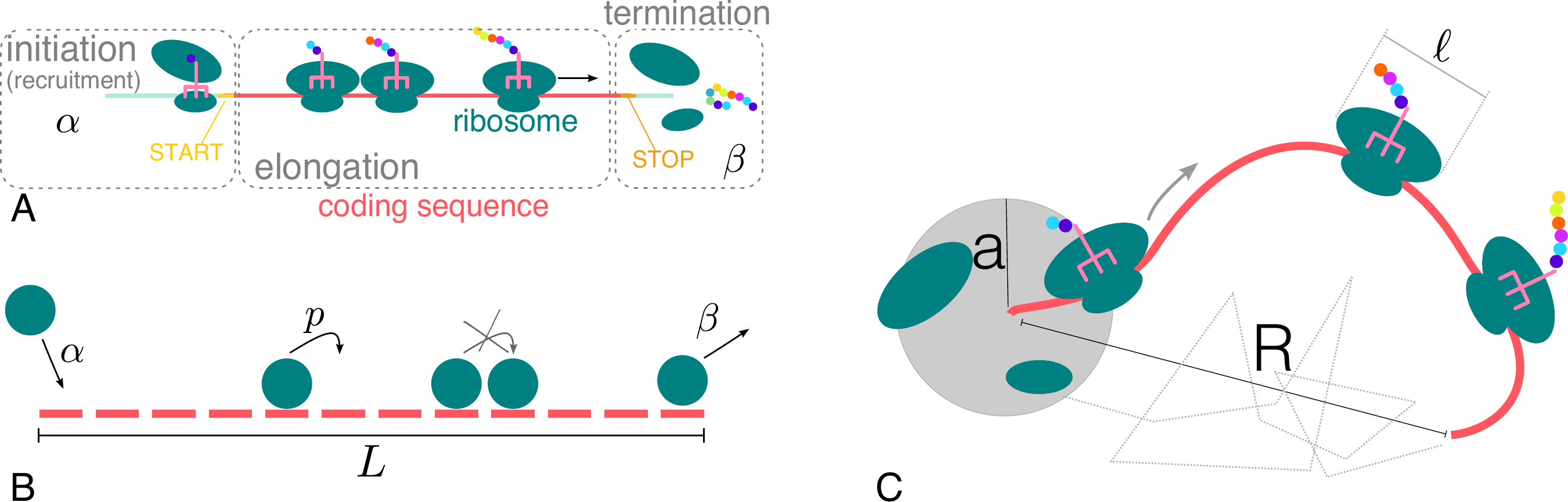}
\caption{Sketch of the translation process and models. The three-steps of the translation process ({\bf A}): initiation (in the model approximated by a one-step process with rate $\alpha$), elongation and termination ($\beta$). In the standard exclusion process ({\bf B}) particles can enter the beginning of the lattice with a rate $\alpha$, move from one site to the next one with rate $p$ (provided that it is not occupied by another particle), and then exit on the last site with rate $\beta$. In this study we consider a more refined version of the model ({\bf C}) in which ribosomes cover $\ell$ sites (codons), advance one site at a time, and the unidimensional lattice is placed in a three-dimensional environment. $R$ represents the end-to-end distance between the 5'and the 3' region, and $a$ is the radius of the reaction volume for initiation. The dashed grey line represents a possible diffusive trajectory of the ribosomal subunits leaving the transcript and being re-absorbed (recycled) in the reaction 
volume around the first site of the lattice. \label{fig::2}}
\end{figure}

The translation of an mRNA is a three-step process, as sketched in Figure~\ref{fig::2}A: during {\it initiation}, the ribosomal subunits are recruited and the full functional ribosome is assembled on the START codon, ready to translate the transcript; then the ribosome proceeds to {\it elongation}, assembling the protein amino acid by amino acid according to the nucleotide sequence of the mRNA; the ribosome eventually detaches when it reaches the STOP codon ({\it termination}). 

We model translating ribosomes as particles moving on a unidimensional discrete track of length $L$ representing the mRNA, as depicted in Figure~\ref{fig::2}B.
In this model particles are injected from one side of the lattice (the 5' end of the mRNA) with a rate $\alpha$, then advance one site (one codon) with rate $p$ only if the arrival site is empty, and are removed at the last site (STOP codon) with rate $\beta$ (Figure~\ref{fig::2}B).
The first step mimics the recruitment of ribosomes (initiation); elongation is given by the dynamics of ribosomes in the bulk; the exit of particles from the last site represents the termination.
MacDonald and coworkers introduced this class of model at the end of the 60's precisely in an attempt to mathematically describe the process of mRNA translation~\cite{macdonald_kinetics_1968}. Since then, under the name of \textit{exclusion process}, this model has been extended and thoroughly studied from a theoretical point of view; it became an emblematic framework in out-of-equilibrium physics~\cite{chou_non-equilibrium_2011}, for which an exact solution is known in the simplest formulation~\cite{blythe_nonequilibrium_2007}.

In the last years, revamped extensions of the exclusion process have appeared in the literature, developed to provide more quantitative models of translation~\cite{mitarai_ribosome_2008, brackley_dynamics_2011, zia_modeling_2011, ciandrini_ribosome_2013}, and many works are nowadays implicitly based on this framework~\cite{reuveni_genome-scale_2011, tian_rapid_2016}. Here we first look into a variant of the exclusion process that considers particles covering $\ell = 10$ sites of the track~\cite{shaw_totally_2003}, as the ribosome footprints cover around 28 nucleotides~\cite{ingolia_genome-wide_2009} (see Fig.~\ref{fig::2}C).  Details of the model and simulations can be found in the Materials and Methods section and in the Supplementary Material.

\subsubsection*{Translation efficiency corresponds to the ribosomal current}
From the analytical solution of the exclusion process or by Monte Carlo simulations we can estimate, as a function of the initiation rate $\alpha$, termination rate $\beta$ and codon elongation rate $p$, the two quantities of interest for the translation process: the ribosomal density $\rho(\alpha, \beta, p)$, defined as the average number of ribosomes $N$ divided by the CDS length $L$, and the ribosomal current $J(\alpha, \beta, p)$, defined as the average number of ribosomes advancing one site in a unit of time. Those quantities can be compared to experimental measurements of polysome profiles and protein production rates. The ribosomal current $J$ corresponds in fact to the protein production rate per mRNA (proteins produced per unit time per mRNA), and we choose to identify it as a better descriptor for the {\it translation efficiency}. 

The same analytical solution also gives the dependence of current and density on the system's parameters. The model's behaviour has been largely studied in the literature as a function of the dimensionless parameters $\bar\alpha \equiv \alpha/p$ and $\bar\beta \equiv \beta/p$, and depending to their values different phases can be observed (see Supplementary Material). 

Since the termination rate $\beta$ is not limiting translation, the translation efficiency and the density do not depend on $\bar\beta$: the system is in the so-called low density phase, in which the density should always be smaller than $\sim 0.076$ consistently with Fig.~\ref{fig::1}. From the analytical solutions one can show that efficiency $J$ and density $\rho$ only depend on $\bar\alpha \equiv \alpha/p$ in this regime (see Supplementary Material). Hence $\rho(\alpha,\beta, p) = \rho(\bar\alpha)$ and $J(\alpha, \beta, p) = J(\bar\alpha)$.\\
%

In order to make the model more realistic and compare it to experimental datasets, we need to determine the initiation rate $\alpha$. The estimation of the translation initiation rate has been previously attempted for a few organisms~\cite{siwiak_comprehensive_2010, ciandrini_ribosome_2013}, and these studies have found a dependence of the initiation rate on the transcript length: the longer the transcript, the weaker the initiation.
Here we propose a model that is able to explain this observation by coupling the translation process, in particular translation initiation, to the three-dimensional conformation of the mRNA. We will show how a feedback mechanism between the ribosomal current leaving the end of the mRNA and the initiation process, which is controlled by the polysome compaction, could induce a length-dependent initiation rate and hence a length-dependent density. Before that, we need to introduce some basic properties of the transcript's spatial conformation.

\subsection*{Transcript end-to-end distance depends on the polysome size}
We consider the transcript as a polymer that assumes different spatial conformations and a characteristic 5'-3' end-to-end distance $R$ (see Figure~\ref{fig::2}C). 
An undecorated mRNA (without ribosomes on it) can be considered as a polymer with a persistence length $l_p \simeq 1$ nm $\simeq 1$ codon~\cite{vanzi_mechanical_2005}, and the average end-to-end distance $R$ can be estimated from basic principles of polymer physics. By assuming an underlying random walk one obtains that the end-to-end distance $R$ depends on the length $L$ of the mRNA as
\begin{equation}
 R=\sqrt{2 L l_p}. 
 \label{EED}
\end{equation}
However, the stiffness produced by the large size $\ell$ of the ribosomes can drastically change the persistence length of the mRNA. We assume that the persistence length of an mRNA depends on the polysome state via an average between $\ell$ (a typical ribosome footprint) and $l_p$ (persistence length of an empty mRNA), weighted by the fraction $f = \rho \ell$ of the transcript covered by ribosomes (at the steady state). After these considerations we write the effective persistence length of the mRNA as
\begin{align}
l_{\textrm{eff}} =& f\ell+(1-f)l_p \notag \\
	    =& \ell^2\rho+(1-\rho \ell)l_p,
\end{align}
which is equal to $l_p$ when the mRNA is empty and reaches $\ell$ when the ribosomal density attains its maximal value $\rho=1/\ell$.
Substituting this value of the effective persistence length into Eq.(\ref{EED}) we obtain the average end-to-end distance of a polysome as a function of the density $\rho$ and the length $L$:
 \begin{equation}
 R =  \sqrt{2 L}\left[\ell^2\rho+(1-\rho \ell)l_p\right]^{1/2}.
 \label{EEDpoly}
\end{equation} 

This way, we have used a coarse-grained model to couple the state of the polysome to its spatial conformation. Intuitively, Eq.(\ref{EEDpoly}) means that a translated transcript with many ribosomes on it will be more stretched (so the distance $R$ between 5' and 3' ends will be larger) compared to a situation with a small ribosomal density or an empty mRNA. 
We have neglected potential formation of secondary structures inside the coding region. Such structures would only slightly decrease the effective length of the sequence (a few codons) and we treat only translationally active transcripts, meaning that moving ribosomes (10-20 codons/s) continuously unfold those structures. 

When initiation is limiting, the density $\rho$ of ribosomes is fixed by $\bar\alpha$, and via Eq.~(\ref{EEDpoly}) we are hence able to determine the dependence of the end-to-end distance as a function of the initiation rate and the length of the transcript, $R(\bar\alpha, L)$.


\subsection*{Initiation can be enhanced by a length-dependent feedback mechanism}

We will consider that the magnitude of the initiation rate $\alpha$ is determined by the concentration $c$ of free ribosomal subunits via a first-order rate equation: $\alpha = \alpha_0 \, c$, with $\alpha_0$ being the initiation rate constant depending, for instance, on the affinity between the ribosome and the 5' UTR binding site on the mRNA. The concentration $c$ is the local concentration of subunits in the reaction volume of radius $a$ around the ribosome binding site at the 5' end of the transcript; we introduce $c_\infty$ as the homogeneous concentration of free subunits far from the transcript. The local concentration $c$ in the reaction volume is affected by the ribosomes terminating translation at the 3' end of a transcript, then diffusing into the reaction volume and contributing to the abundance of free ribosomal subunits in this volume. The contribution  $\delta_R$ to the local concentration $c$ due to this feedback mechanism depends on the end-to-end distance $R$ previously calculated in Eq.~(\ref{EEDpoly}). It can be shown for different organisms, considering typical values of transcript numbers and cytoplasm volume, that the average end-to-end distance is smaller than the average separation between transcripts. This corroborates the intrinsic assumption that the translating processes of distinct mRNAs do not interfere with each other.
Thus each individual mRNA can be thought of as a sink-source system for ribosomes (the ribosome binding site representing the sink and the ribosome termination site representing the source) immersed in an environment with a constant background ribosomal concentration $c_\infty$. Hence, the local ribosome concentration around the ribosome binding site can be written as $c(R) = c_\infty + \delta_R$. 

\begin{figure}[h!b]
\centering
\includegraphics[width=0.5\linewidth]{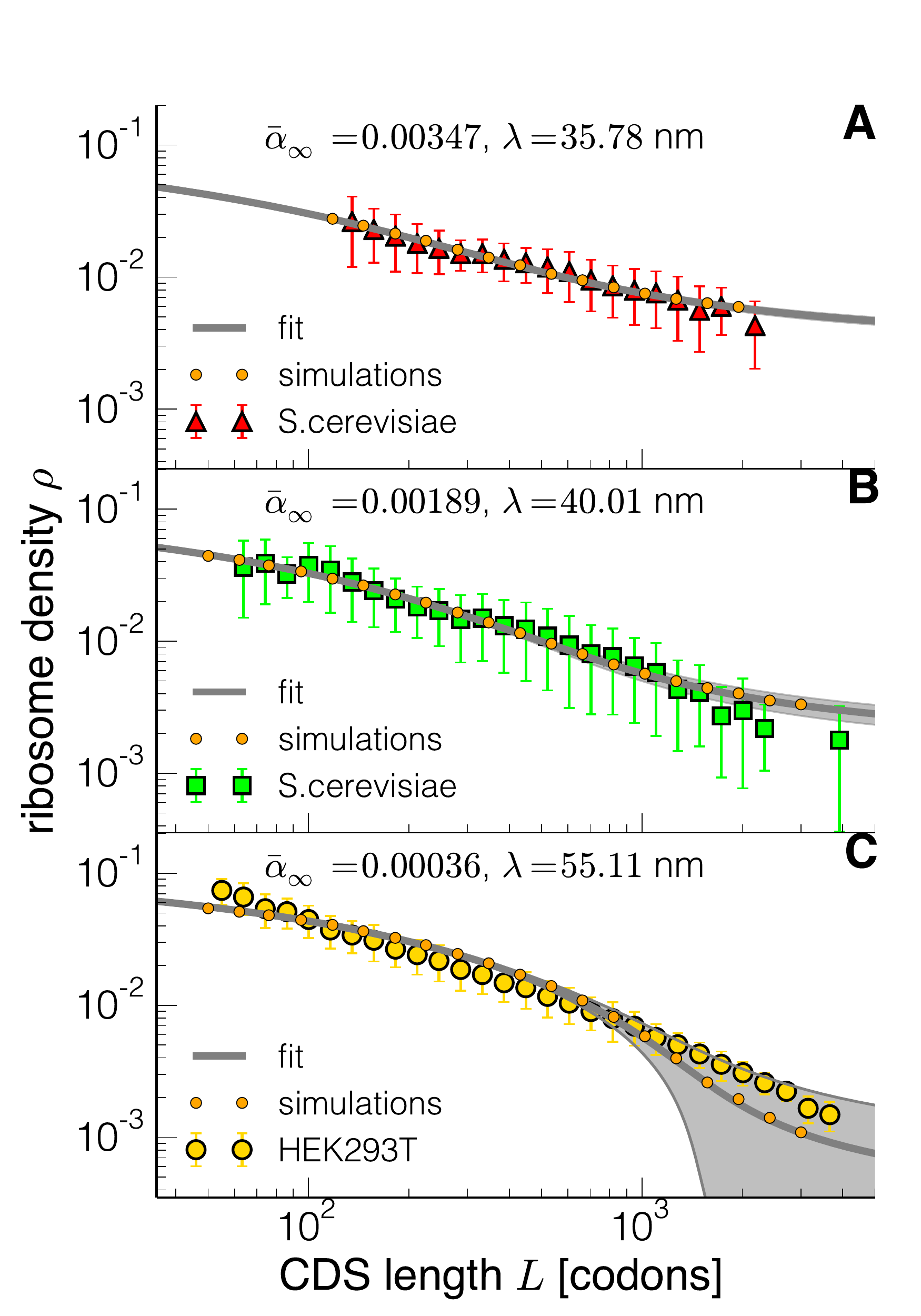}
\caption{Comparison between theory and experimental ribosome densities in yeast ({\bf A-B})\cite{arava_genome-wide_2003, mackay_gene_2004} and human embryonic kidney cells ({\bf C})\cite{hendrickson_concordant_2009}. The symbols and datasets correspond to the ones of Figure~(\ref{fig::1}). The grey lines represent the best fit of the model (the parameter values are written in each panel), while the shadow areas correspond to the regions spanned within the margin of error of the estimated $\bar\alpha_\infty$. Orange circles are the outcome of stochastic simulations used to test the numerical solution of the equation using the parameters obtained from the best fit. 
\label{fig::4}}
\end{figure}

By considering ribosomes as particles performing free diffusion when they are not bound to the mRNA (see the Supplementary Material) we can obtain a mathematical expression of the initiation rate as a function of the system's parameters. It can be shown (see Supplementary Material) that the increase $\delta_R$ due to the source at the end of transcript is $\delta_R \sim J/DR$, where $D$ is the diffusion coefficient of ribosomes ($\sim 0.04\mu$m$^2$/s~\cite{bakashi_2012}). Considering that the protein production rate per transcript is $J\sim 0.1-10$ proteins/s~\cite{ciandrini_ribosome_2013, Bremer1987}, and that in {\it E.coli} $c_\infty\sim 0.5-1\cdot 10^3$ free ribosomes$/\mu m^{3}$~\cite{Arkin1633}, for a typical mRNA of length $\sim300$ codons this back-of-the envelope calculation leads to $\delta_R/c_\infty \sim 0.1-10$, meaning that the concentration increase is at least comparable with the bulk ribosomal concentration and the mechanism proposed is relevant in the biological regime. 

Regarding translation as a steady state process with ribosomal density and current values given by $\rho$ and $J$ respectively, we find the initiation parameter $\bar\alpha$ (see Supplementary Material for a complete derivation):
\begin{equation}
 \bar\alpha =  \bar\alpha_0 (c_\infty + \delta_R) =  \bar\alpha_\infty + \lambda \frac{J(\bar\alpha)}{R(\bar\alpha,L)}, 
 \label{initiation}
\end{equation}
where we have emphasised the dependence of the ribosomal current $J$ on $\bar\alpha$. 
Similarly, the end-to-end distance $R$ also depends on $\bar\alpha$ and $L$ as shown in Eq.~(\ref{EEDpoly}).
We highlight that the parameter $\bar\alpha_0$ and thus $ \bar\alpha_\infty$ and $\lambda$ depend on the binding between the ribosome and the mRNA, which is supposed to be mainly regulated by mRNA secondary structures. We will consider the parameters of the model to be independent on the transcript length. This is justified by a weak correlation (Pearson $r = - 0.01$, p-value $0.5$)~\cite{ringer_folding_2005} between free energies of secondary structures in the 5'UTRs and the transcript length $L$ (see also Figure S6 in the Supplementary Material). 
The parameter $\bar\alpha_\infty$ is adimensional and $\alpha_\infty = p \bar\alpha_\infty$ represents the initiation rate without the feedback mechanism between the current and the initiation, or equivalently when the end-to-end distance $R$ is large enough to make this mechanism negligible. The parameter $\lambda$ characterises the strength of the feedback. It has the dimensions of a length and it corresponds to the typical separation between 5' and 3' below which the feedback mechanism becomes relevant. We measure this parameter in units of codon length, which roughly corresponds to $1$ nm. 

Consequently, the current of ribosomes leaving the end of a transcript increases the concentration of ribosomal subunits around their binding region; through modulation of the mRNA stiffness due to the ribosome load, this feedback leads to initiation rates that are strongly length-dependent. Equation (\ref{initiation}) is an implicit equation that can be numerically solved to obtain the initiation rate, and thus the density $\rho(\bar\alpha)$ and the current as a function of $L$. To check the validity of our analytical calculations, we also developed a simulation scheme that allows us to fix the initiation rate via a self-consistent method (see Methods section).

Although Eq. (\ref{initiation}) considers the ribosome as a single diffusing particle, we can explicitly consider the diffusion of the two ribosomal subunits. This would generate a dependence on $R^2$ instead of $R$ in Eq. (\ref{initiation}). However, the qualitative behaviour of our results does not significantly change and for the sake of simplicity we decided to present the outcomes of the theory described by Eq. (\ref{initiation}). We include the analysis of this more refined model in the Supplementary Material (see also Figure S9 and S12).

\subsection*{Experimental density-length dependence emerges from initiation enhanced effects}
We then compare the outcome of the model to experimental measurements of ribosome densities. The result of this analysis is shown in Figure~\ref{fig::4}: the predicted ribosome density is quantitatively comparable to the experimental quantification, and our mechanistic model based on basic physical principles is able to capture the length dependence of the ribosome load. Our theory can therefore explain the observed length dependence of the translational properties.

We fit the expression of  $\rho(\bar\alpha)$ (continuous lines in Figure~\ref{fig::4}) and obtain the two parameters $\bar\alpha_\infty$ and $\lambda$ for three available datasets (two yeast datasets~\cite{arava_genome-wide_2003, mackay_gene_2004} and a human embryonic kidney cells dataset~\cite{hendrickson_concordant_2009}), then check the accuracy of the solution with the stochastic simulation scheme developed as explained in the Material and Methods section. 
Taking the standard errors of the parameter estimation there is no significant deviation between the data and the fitted model(details can be found in the Materials and Methods). However, for the last set, there is a stronger dependence of the solution on the parameter $\bar\alpha_\infty$: shaded regions in Figure~\ref{fig::4} represent the solution considering the standard error of this parameter. This could explain the slight deviation between theory and experimental data for large mRNAs. We also emphasise that the parameter $\bar\alpha_\infty$, depending on the global availability of ribosomes, is supposed to be the most affected by experimental variations (for instance by growth rate dependence or cell cycle stage). The estimation of $\bar\alpha_\infty$ is more accurate when we take into account the diffusion of the two subunits (see Supplementary Material).

Our simulations also allow us to extract the amount of ribosomes bound to a transcript, from which we can extract the monosome:polysome ratio, which is also subject to length-effects (see Supplementary Material). By increasing the CDS length we observe a reduction of the monosome:polysome ratio following a power-law like behaviour.
A recent work~\cite{heyer_redefining_2016} has identified, by merging polysome and ribosome profiling techniques, the amount of active monosomes, i.e. mRNAs with only one ribosome elongating the protein. Consistent with our findings, the monosome:polysome ratio also shows signs of a marked anti-correlation with the mRNA length.

\subsection*{Ribosome drop-off cannot alone be responsible for the density-length dependence}
\begin{figure}[h!t]
\centering
\includegraphics[width=0.5\linewidth]{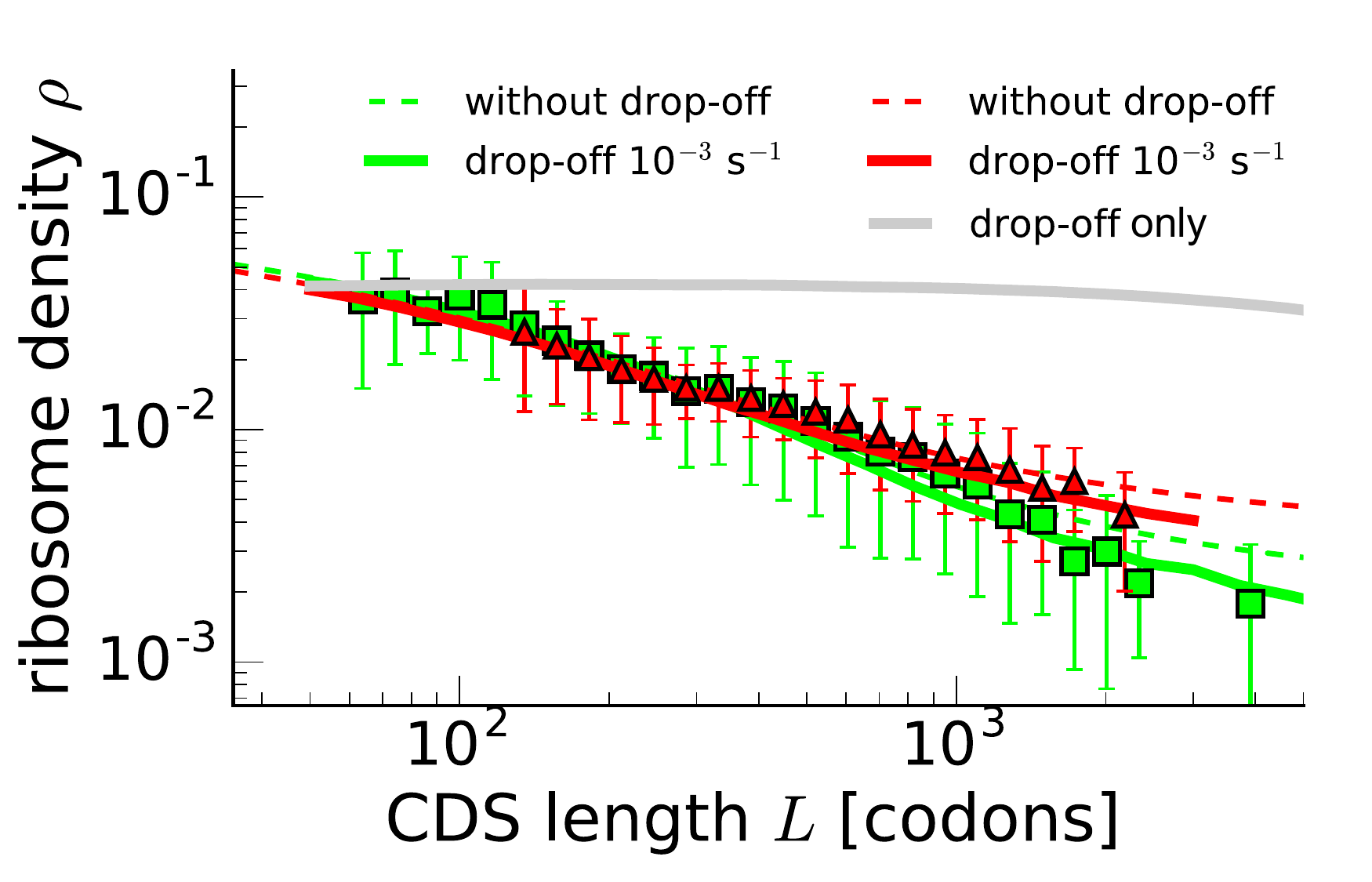}
\caption{Model with ribosome drop-off. Green lines correspond to the Arava dataset~\cite{arava_genome-wide_2003}, while red ones correspond to the Mackay dataset~\cite{mackay_gene_2004} Symbols of experimental points in yeast correspond to the ones of Figure~\ref{fig::1}, while dashed lines represent the solutions of the model described in the previous section with the same parameters used in panels A and B of Figure~\ref{fig::4}. The continuous lines are the outcome of simulations of our model including ribosome drop-off at a rate of $10^{-3}$ s$^{-1}$, and the grey line shows the outcome of the simulations with drop-off only.
\label{fig::dropoff}}
\end{figure}

In this section we study the consequences of ribosome drop-off~\cite{sin_quantitative_2016, bonnin_novel_2017}, thus far neglected in our model, on the observed density-length dependence. In order to do that, we performed simulations including ribosome drop-off at a rate $\delta = 10^{-3}$ s$^{-1}$. This is justified by the estimated drop-off rates of the order of $10^{-4}$/codon~\cite{sin_quantitative_2016, bonnin_novel_2017}, and by the codon elongation rates we considered in this paper of the order of $10$ codons s$^{-1}$.
Figure~\ref{fig::dropoff} shows that the simulations of the process with drop-off (full lines) do not largely differ from the model without drop-off (dashed lines), and the deviations starts to become relevant for large sequences (order $10^4$ codons). For such lengths the extended model is actually reproducing the experimental behaviour even better than the model without drop-off.

To further exclude the possibility that the length dependence rises from ribosome prematurely leaving the transcript, we simulated ribosome drop-off occurring during the translation process without the feedback mechanism that we propose (basically, we simulated a standard TASEP with particle detachment rate $\delta$). This is represented by the grey line in Figure~\ref{fig::dropoff}. We were not able to obtain any length dependence for biologically relevant values of the drop-off rate (see Supplementary Figure S10), and we can therefore conclude that the behaviour observed in Figure~\ref{fig::1} cannot originate by ribosomal drop-off alone.

\subsection*{mRNA circularisation does not change the phenomenology of the model}
\begin{figure}[t!h]
\centering
\includegraphics[width=0.5\linewidth]{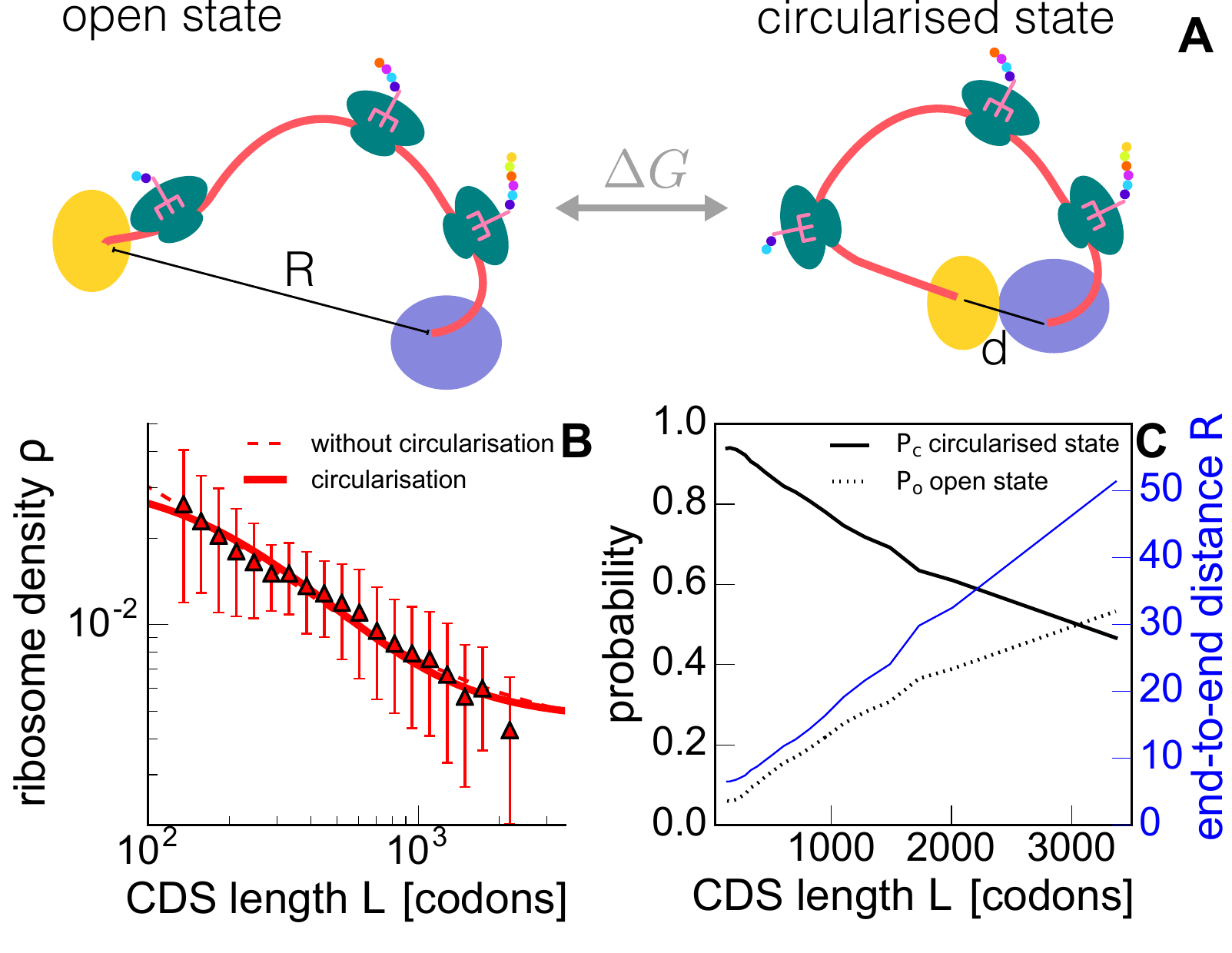}
\caption{Model with mRNA circularisation. Sketches of the two possible mRNA conformations, open and circularised, whose transitions depend on the free energy gap $\Delta G$ ({\bf A}). In blue and yellow we have represented the interacting proteins (e.g. PABP and eIF4F) bound at the 3' and 5' ends; the black line is the end-to-end distance that is equal to $d$ in the circularised state. We have fixed $d=5$ nm in our calculation. Ribosome density computed taking into account mRNA circularisation (continuous line) is then compared to experimental data (triangles, cfr. symbols used in Figure~\ref{fig::1}) and the previous model neglecting circularisation (dashed line) ({\bf B}). The fitted parameters are $\bar\alpha_\infty = (4.7 \pm 0.6)\ 10^{-3}$ s$^{-1}$, $\lambda = 7.0 \pm 0.6$ nm and $\epsilon = -8.3 \pm 0.4$ ($k_B T$). End-to-end distance (blue curve, right axis) and calculated probabilities $P_c$ and $P_o = 1- P_c$ as a function of the CDS length $L$ ({\bf C}).}
\label{fig::circ}
\end{figure}

The 5' and 3' end of eukaryotic mRNAs interact with each other via protein-protein interaction, for instance between the Poly(A)-binding protein PAPB and the initiation factor eIF4F bound at the 5' cap; this coupling is believed to induce the formation of transcripts with circular structures~\cite{Wells1998135}. However, depending on the energies at play, the transcript dynamically switches between an {\it open}, linear state, and a {\it circularised} state (see Figure~\ref{fig::circ}A). When in the circularised state, the end-to-end distance $R$ will be only of a few nanometers (the order of magnitude of the two molecular partners supposedly involved in this interactions). We will denote as $d$ the distance between 5' and 3' in the circularised state. 
The mRNA is found in its circularised state with a given probability $P_c$, and in an open state with probability $P_o = 1- P_c$, with a difference in free energies between the two states $\Delta G = G_c - G_o$.
The free energy difference depends on the physical parameters of the model such as $l_\textrm{eff}$, $L$, $\epsilon$ and $d$ (as it can be seen in Eq.~(S19)), and how $\Delta G$ increases as a function of these parameter is briefly discussed in the Supplementary Material.

To find how the average end-to-end distance is affected by this interaction we weight its value in the circularised and open state with their respective probabilities:
\begin{equation}
	R_\textrm{circ} =  P_o R + P_c d \,.
	\label{R_circ}
\end{equation}

If $P_o \approx 1$ then our feedback model well approximates the mRNA translation process. 
While in prokaryotes we can likely assume $P_o = 1$, this is probably an oversimplification for eukaryotes. In order to consider  transcript circularisation we have now to compute how $P_o$ depends on the CDS length $L$ and on the interaction energy $\epsilon$ (in $k_B T$ units) between the two ends. The intuitive explanation we used before to determine differences in local concentrations of ribosomes close to the 5' end can be reproduced here to compute $P_c$. For a fixed $\epsilon$ and a short mRNA, we expect to find circularised transcripts with a probability $P_c$ close to one; in contrast, very large transcripts should be hardly found in the circularised state. This length dependence will also contribute to the ribosome density behaviour observed in experiments (Fig.~\ref{fig::1}). We computed $P_c$ as a function of $L$, $\ell_\textrm{eff}$ and $\epsilon$, which turns out to be:
\begin{equation}
P_c = \cfrac{1}{1+e^{\Delta G/k_B T}}  =
       \cfrac{1}{1+\left[\left(\cfrac{l_\textrm{eff}L}{d^2}\right)^{\frac{3}{2}}\sqrt{\cfrac{4\pi}{3}}-1\right]e^{\left(\cfrac{2\pi^2 l_\textrm{eff}}{L}+\epsilon\right)}} \,.
       \label{eq::Pc}
\end{equation}

The details of the calculation of $P_c$ can be found in the Supplementary Material. Here $\epsilon$ is considered as a fitting parameter.
By inserting Eq.(\ref{eq::Pc}) in Eq.(\ref{R_circ}) and computing the end-to-end distance to be plugged in Eq.~(\ref{initiation}) we eventually find, now as a function of $\bar\alpha_\infty$, $\lambda$ and $\epsilon$, how the initiation rate is affected by the concentration increase of ribosomes in the 5' reaction volume with also considering transcript circularisation. We have then fitted $\rho(\bar\alpha)$ to the dataset we have previously used, and the outcome is shown in Figure~\ref{fig::circ}B (fitting values and other datasets can be found in the Supplementary Material). We did not find a significative difference from the best fit of the simpler model previously introduced (dashed line in Figure~\ref{fig::circ}B). In Figure~\ref{fig::circ}C we show how the probabilities of finding a circularised or open mRNA depend on the transcript length, together with the end-to-end distance $R$. The results agree with our intuitive explanation.

\subsection*{Length-dependent competition for resources}
In this section we speculate on some potential applications of our model related to bacterial growth laws~\cite{scott_interdependence_2010, Dai2016}. Specifically, we will show that cells can adjust, based on a length-discriminatory mechanism, their relative expression of genes at different ribosome concentrations.  
This result from our model leads us to predict and theorise a new regulation mechanism for gene expression.

\begin{figure*}[h!t]
\centering
\includegraphics[width=0.65\linewidth]{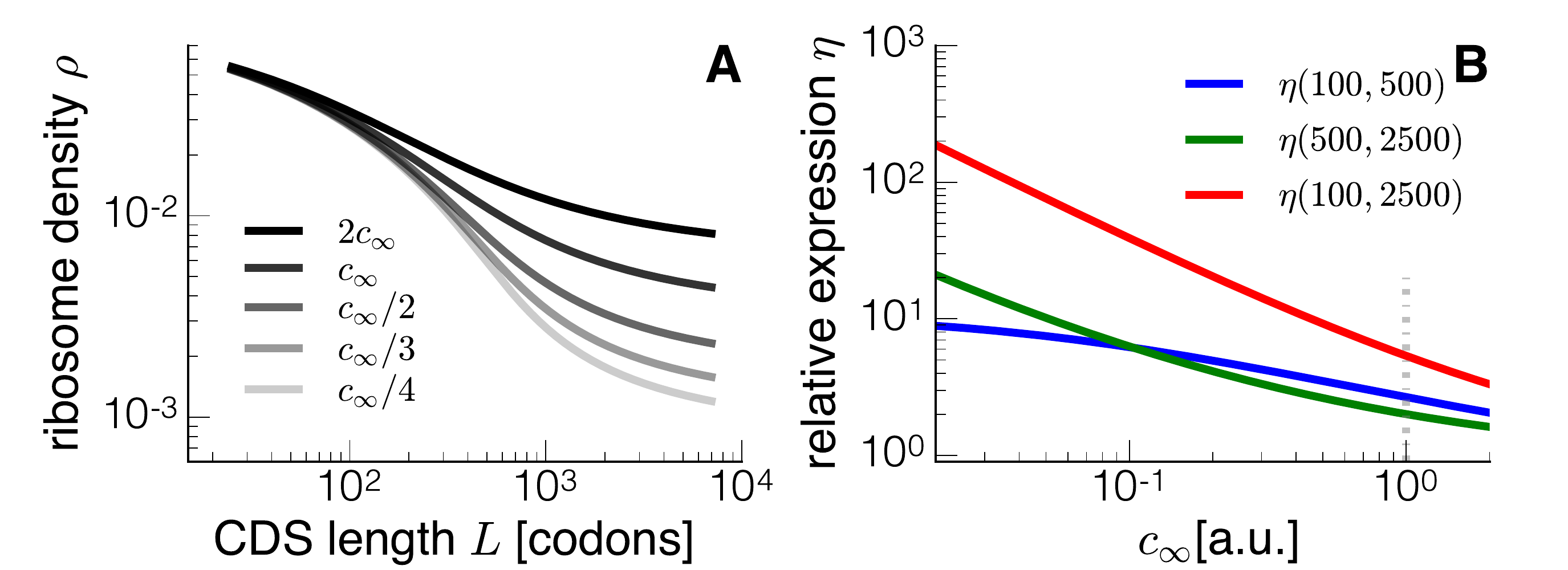}
\caption{Effects of competition for resources (ribosomes) on the protein production rate. The ribosome density depends on the overall ribosome concentration $c_\infty$. We show the ribosome density as a function of the transcript length for different concentrations of available ribosomes $c_\infty$ ({\bf A}). The curve denoted with $c_\infty$ in the legend is built starting from the same parameters of Figure~\ref{fig::4}A. We change $c_\infty$ as described in the legend for the other curves.
 Short transcripts are less affected by changes in $c_\infty$, as we also show in ({\bf B}), where we plot the relative expression of transcripts $\eta$ (defined in the text) as a function of $c_\infty$. We used transcript with three different lengths (here $L = 100, 500$ and $2500$). According to these results, ribosomal proteins that are short should be relatively more expressed under high ribosome competition regimes compared to other types of proteins.
 \label{fig::5}}
\end{figure*}

We observe that changes in ribosome densities (or in translation efficiency) induced by changes in the ribosomal pool are conditional on the mRNA length (Figure~\ref{fig::5}A). Short transcripts are less affected by the amount of available ribosomes compared to long ones, suggesting a possible mechanism to regulate the relative protein production rate at different growth rates based on transcript length only. As a proof of principle, in Figure~\ref{fig::5}B we plot the relative expression $\eta (L_1, L_2) = J_{L_1}/J_{L_2}$ between the translation efficiencies of transcripts with lengths $L_1$ and $L_2$. When ribosomes are limiting, short transcripts are relatively more translated than long ones.

This behaviour can be intuitively explained. The main contribution to the initiation of long mRNAs is the concentration of free ribosomes $c_\infty$, meaning that long mRNAs are largely influenced by changes in the ribosomal pool. Short transcripts instead can more efficiently take advantage of the length dependent contribution $\delta_R$ to initiation. 

This constitutes a potential mechanism for regulating the relative expression between short and long genes. Ribosomal proteins are composed only of a few dozens of amino-acids and, as a consequence, our theory predicts that at least qualitatively ribosomal proteins should be proportionately more efficiently translated than longer proteins under strong ribosome competition regimes, i.e. low growth rates. This difference should decrease when ribosomal resources get less tight, and in the limit of infinite ribosomal resources, $\eta$ should tend to $1$ since the length dependence becomes less and less relevant.

Our model suggests a translational mechanism to relatively over express short proteins (e.g. ribosomal proteins) at the cost of longer ones.

\section*{Discussion}

Many aspects of translation are still puzzling researchers. Theoretical approaches propose designing principles for modulating the translation efficiency at the level of initiation~\cite{salis_automated_2009} and of elongation~\cite{ciandrini_ribosome_2013}, mainly based on the role of RNA secondary structures, codon bias or amino acid properties. However, when tested on synthetic constructs, the hypotheses underlying the theoretical models are often contradicted~\cite{li_how_2015}, so that the identification of transcript-dependent determinants of translation efficiency is still debated in the literature. In this work we have identified and studied another factor modulating the translation efficiency: the length of the transcript. 

According to our results, the proximity of the 3' end to the ribosomal recruitment site of the mRNA could induce a feedback in the translation process that would favour the translation of short transcripts over long ones, as has been shown by experiments in the last decades.
We connected the emergence of the ribosome density-mRNA length dependence shown in Figure~\ref{fig::1} to a mechanistic model built on basic physical principles and on the properties of the translation process. Our theory then establishes a link between densities and mRNA lengths and explains experimental data with an excellent agreement without invoking an evolutionary selection of genes based on their length. As a matter of fact, a selection process towards short efficient genes to improve cellular fitness could also be conjectured~\cite{castillo-davis_selection_2002, eisenberg_human_2003}. Without direct experimental observation we cannot rule out this hypothesis, although this would not explain the same behaviour observed for different organisms (experimental data in Figure~\ref{fig::1} seems to collapse to a unique universal curve). Moreover, the poor correlation between free energy of secondary structures at the 5' end of an mRNA and CDS length is a signature that binding sites are not significantly 
weaker for long genes, as it would assume an evolutionary argument. Instead our theory predicts, as an outcome, that short genes have larger initiation rates compared to long ones. A recent publication by Li {\it et al.}~\cite{Li2017} confirms that transcript length is a main determinant of translation.

Experimental results in fact suggest that translation initiation is also dependent on mRNA length~\cite{arava_genome-wide_2003,arava_compaction_2009}. Since the ribosome recruitment rate must depend on the local concentration of ribosomal subunits around the 5'UTR, we conjecture that the local concentration is modulated by the CDS length via a feedback mechanism coupling protein synthesis and initiation rates. We can roughly name this process ``recycling'', as subunits terminating translation will contribute to the increase of the local concentration $\delta_R$ and thus be more easily re-used as sketched in Figure~\ref{fig::2}. This coarse-grained physical model has allowed us to reproduce the scaling behaviour of experimental ribosome densities (Figure~\ref{fig::4}) and it constitutes, to our understating, the first quantitative explanation of how translational features are affected by the transcript length. 
Previous works have studied the effect of particle recycling~\cite{chou_ribosome_2003, sharma_stochastic_2011, margaliot_ribosome_2013, marshall_ribosome_2014} but, with the exception of Chou (2003)\cite{chou_ribosome_2003}, they do not explicitly compute how the recycling term is regulated by the end-to-end distance $R$.

The compaction of the transcript (here characterised by the end-to-end distance), also depends on its polysome state. Intuitively, an mRNA with many translating ribosomes will be more stretched than an empty mRNA. We captured this feature by introducing a ribosome density dependence on the end-to-end distance through Eq.~(\ref{EEDpoly}). We emphasise that this is the simplest choice for coupling elongation properties and the three-dimensional conformation of the mRNA, and one could introduce more complicated relationships linking end-to-end distance, ribosome density and elongation rate; here we wanted to show that, as a proof of principle, by a feedback mechanism enhancing initiation we can reproduce experimental data very well. 

We have also studied how the results change by explicitly considering the diffusion of the two ribosomal subunits, and we found no significant change (see Supplementary Material).   
To make the model more realistic we considered two further extensions of the model. We considered (i) ribosome drop-off and (ii) mRNA circularisation. The former brought an improvement in the comparison between data and theory for large transcripts  only (see Figure~\ref{fig::dropoff}), while the inclusion of the latter did not lead to a particular phenomenological change of the model's outcomes (Figure~\ref{fig::circ}). This suggests that ribosome recycling, as considered in the basic model, is the fundamental element originating the length-dependence translation.

We have then speculated on how the length could be exploited to create differences in the relative expression of genes at different growth rates, here used as measure of the free ribosomes abundances. When resources are constrained, i.e. when the amount of free ribosomes $c_\infty$ is small, competition for resources might become relevant~\cite{greulich_mixed_2012, shah_rate_2013, raveh_a-model_2016} and our theory predicts that the length-dependent term of the initiation rate dominates the process. In other words, when ribosomes are strongly limiting, ribosome recruitment is mainly due to recycled ribosomes, meaning that short transcripts can better capitalise the resources. This mechanisms could also be a way to translationally favour the production of ribosomal proteins (which are short) in a scenario of deficiency of ribosomes.
In order to formulate this hypothesis, we neglect known mechanisms responsible for ribosome biogenesis, a complex process that is beyond the scope of our work. 
Our conjecture has then to be interpreted in the perspective of ribosomal concentrations fixed by a certain amount of ribosome production (established, for instance, by the richness of the growth medium): for a given concentration of ribosomes we make strong predictions on how the ribosomal pool should be partitioned among the different transcripts with just a length-discrimination mechanism.

The model could be further extended to consider translation of bacterial operons: in this case, in fact, one transcript is composed of different sinks (ribosome binding sites) and sources (stop codons) of ribosomes, while here we have discussed the case of a transcript translating a single gene (with one ribosome binding site and one stop codons). Having an operon translating different genes will increase the complexity of the feedback term, and it could in principle create counter-intuitive phenomenologies.

Our findings are compared to experimental ribosomal densities, and our framework can quantitatively reproduce the measurements.
We have used our model to estimate the protein production rates of synonymous genes, and the method was successful (see Figure S13). In this work we have used our model to predict the ribosome-length dependence, i.e. we have emphasised the dependence of Eq.~(\ref{initiation}) on $L$, but our theory predicts that there is a feedback between elongation (codon usage) and initiation, that we have exploited in Figure S13.
However, to further test the model and the relevance of the regulatory mechanism we propose, 
 it will be necessary to make fusions of a reporter gene with peptides of variable lengths, and then measure ribosome density or translation efficiency with the aim of experimentally reproducing Figure~\ref{fig::4} in a controlled manner. However, one should pay attention to the changes in mRNA degradation, translation elongation and initiation induced by the added nucleotides coding the fused peptides. Figure~\ref{fig::5}B also constitutes a good way to test our hypotheses. One of our predictions is the relative change of expression of short/long transcripts when changing the cellular growth rate. This could be obtained by concurrently expressing two reporter genes of different lengths, and measuring their relative expression at different growth rates obtained by changing growth medium or by different antibiotics.

\section*{Methods}

\label{sec::mm}

\subsection*{The exclusion process}
We base our model on the exclusion process, which is also introduced in the section Results and in Figure~\ref{fig::2}B. More accurately, this model is known in the literature as TASEP: Totally Asymmetric Simple Exclusion Process, for which nowadays there exists a plethora of extensions applied in many different fields, from vehicular traffic to intracellular transport~\cite{chou_non-equilibrium_2011}.
Each site of the track in Figure~\ref{fig::2}B corresponds to a codon, and the particles can advance from site to site -provided that the next site is not occupied by another particle- mimicking the elongation process. 
We give a thorough description of the exclusion process and the known results in the Supplementary Material. 

To simulate the dynamics of the exclusion process we used a kinetic Gillespie-like Monte Carlo as used in Ciandrini {\it et al.}~\cite{ciandrini_ribosome_2013}.

\subsection*{Fitting and numerical solutions}
We substitute the expression for $\bar\alpha$, Eq.~(\ref{initiation}), in the equation for the density $\rho(\bar\alpha)$ in the low density phase of the $\ell$-TASEP (see Supplementary Material). The current $J$ is given by the $J(\rho)$ correction in the $\ell$-TASEP and the end-to-end distance is $R$ found in Eq.(\ref{EEDpoly}). Thus, we obtain an implicit equation $\rho = \rho(\bar\alpha_\infty, \lambda, L)$ that can be numerically solved for each set of variables $\{\bar\alpha_\infty, \lambda, L\}$ and used to fit the experimental data $\rho_\textrm{exp}(L)$ to obtain the parameters $\bar\alpha_\infty$ and $\lambda$ for each dataset, and their standard errors (see Supplementary Material).

We have used built-in functions of Mathematica \cite{math} to obtain numerical solutions for the density and currents and to fit the three datasets used in this study.

\subsection*{Density and current via a self-consistent simulation scheme}

Equation (\ref{initiation}) allows us to obtain, via simulation, the values for $\rho$ for different values of $L$, taking into account the feedback mechanism coupling protein synthesis and initiation rate and finite size effects (the later intrinsic to the numerical simulations). We obtain this with the following self-consistent method:

 \begin{itemize}
	 \item[(i)] We initialise the system with an arbitrary value of $\alpha = \alpha^{(0)}$, let the system evolve until the steady state is reached and then evaluate the current $J^{(0)}$ and the density $\rho^{(0)}$;
	 \item[(ii)] Compute $R$ as in equation (\ref{EEDpoly}) and update $\alpha = \alpha^{(1)}$ according to equation (\ref{initiation}) with $J^{(0)}$ and $\rho^{(0)}$ computed in (i);
	 \item[(iii)] Repeat the previous points for several iterations until $|\alpha^{(i)}-\alpha^{(i-1)}|/\alpha^{(i)} < 0.01$ (in general less than 10 iterations are needed to make the algorithm converge);
	 \item[(iv)] The final value of $\alpha$ is then used to obtain the final densities and currents.
 \end{itemize}
 
With this iteration process, for a given choice of the parameters $\alpha_{\infty}$ and $\lambda$, we can obtain the steady state density and current, which vary with the length $L$ of the transcript, due to the joint contribution of recycling and finite-size effects.

This self-consistent method allows us to simulate the system without explicitly considering, thanks to Equation (\ref{initiation}), the diffusion of particles when they are not bound to the lattice and the dynamics of the mRNA.

\subsection*{Choice of datasets}
We restricted our analysis to measures made by sucrose gradient methods. We are aware that a more recent technique like ribosome profiling\cite{ingolia_genome-wide_2009} would provide ribosome densities with codon resolution, but this method does not provide an {\it absolute} ribosome density (see definition below). The length-correlation has been shown to hold in ribosome profiling experiments\cite{ingolia_genome-wide_2009}. Instead of assuming arbitrary normalisation of ribosome footprints to match our theory, we analysed absolute ribosome densities that are available in the literature.
For the Mackay {\it et al.} dataset\cite{mackay_gene_2004} we used the {\it reliable} subset of data. 

\subsection*{Definitions}
We define the ribosome density to be the number $N$ of translating ribosomes divided by the length $L$ (expressed in number of codons) of the CDS. We embrace this definition instead of alternative ones (ribosomes per 100 or 1000 nucleotides) for practical reasons. Thus, the density $\rho \equiv N/L$ is expressed in ribosomes per codons, and it can be thought of as the probability of a codon being covered by the the A-site of the ribosome (i.e., a codon being translated). For steric reasons, this density is bound by 1/$\ell$, where $\ell$ is the length of the ribosome footprint (in codons). For instance, $\ell\sim10$ in \textit{S. cerevisiae}.

\bibliography{length_dependence_}

\section*{Acknowledgements}
The authors would like to acknowledge the funding provided by the European Union Seventh Framework Programme [FP7/2007-2013] (NICHE; grant agreement 289384) (LDF). LDF also acknowledges the funding provided by the S\~{a}o Paulo Research Foundation (FAPESP - grant \#2015/26989-4). AM was partially funded by the UK Biotechnology and Biological Research Council (BBSRC), through grant BB/N015711/1. LC would like to acknowledge Maria Carmen Romano, Jean Hausser, Marco Cosentino Lagomarsino, Jean-Charles Walter and Norbert Kern for early discussions on this work, and the CNRS for having granted him a ``demi-d\'{e}l\'{e}gation'' (2017-18). We would like to dedicate this work in memory of Maxime Clusel and Vladimir Lorman.

\section*{Author contributions statement}
L.C. and A.M. conceived and discussed the theory, L.C. and L.D.F. performed the simulations and analysed the data, L.C., L.D.F. and A.M. wrote the manuscript text and supplementary information.  All authors reviewed the manuscript. 

\section*{Additional information}
\textbf{Competing financial interests}. None declared.

%
%

\end{document}